\documentclass[preprint,aps]{revtex4}
\usepackage{ucs}
\usepackage[utf8x]{inputenc}
\usepackage{amsmath}
\usepackage{subfigure}
\usepackage{float}
\usepackage{rotating}
\usepackage[hang,bf]{caption}
\usepackage{hyperref}

\begin{document}
\title{Restricted Open-Shell Kohn-Sham Theory IV:
Expressions for N Unpaired Electrons}

\author{Marius Schulte}
\affiliation{Department Chemie und Biochemie, Butenandtstr. 11, 81377 M{\"u}nchen, Germany}
\author{Irmgard Frank}
\affiliation{Institut f{\"u}r Physikalische Chemie und Elektrochemie, Leibniz
Universit{\"a}t Hannover, Callinstr. 3A, 30167 Hannover, Germany}

\begin{abstract}
We present an energy expression
for restricted open-shell Kohn-Sham theory
for N unpaired electrons 
and single-electron operators for all multiplets
formed from up to five unpaired electrons.
It is shown that it is possible to derive an explicit
energy expression for all low-spin multiplets of systems that exhibit
neither radial nor cylindrical symmetry.
\end{abstract}

\maketitle

\section{Introduction}
Restricted open-shell theory \cite{Roothaan60,Davidson73}
and in particular restricted open-shell
Kohn-Sham theory (ROKS) \cite{Frank98,Grimm2003,Friedrichs2008} has gained
renewed interest in recent years due to its
application in the simulation of photoreactions \cite{Nonnenberg2003,
Roehrig2004,Grimm2005,Nonnenberg2005,Nonnenberg2006,Frank2007}.
While the concept of restricted open-shell theory is seemingly
simple, only part of the states of interest are accessible in present
implementations.
For the high-spin case (multiplicity = number of unpaired electrons + 1)
the situation is simple: only one Slater determinant is needed for the
description of the wavefunction. The most trivial case is the
one with but one unpaired electron which gives rise to a doublet
that is described by a single determinant in a straightforward way.
For most single-determinant cases,
results very similar to the unrestricted theory are obtained
and the use of the restricted theory is not necessary.
For low-spin cases the situation is much more involved and
from a purely mathematical view one might doubt if these cases
can be analysed at all.
One of the general criticisms concerning open-shell theory has already been
addressed by Roothaan in his original
restricted open-shell Hartree-Fock (ROHF) paper \cite{Roothaan60}:
He showed that although orbital-dependent operators are obtained,
the equations can be augmented to a Hermitean
formulation by the use of projection operators. This guarantees 
diagonalization to real eigenvalues.
For low-spin cases there is the additional problem of 
orbital rotations leading to unphysical localization
which has also been noted as sudden polarization in related
approaches \cite{Salem76}, and there is a lot of confusion
in the literature due to the difficulty in correctly
interpreting and avoiding this phenomenon of unphysical orbital localization. 
Undesired rotations can in be avoided 
by exploiting spatial symmetry or
by state averaging; general self-consistent
solutions for the restricted open-shell problem
are difficult to achieve \cite{Davidson76}. In recent years we have shown
for the case of restricted open-shell Kohn-Sham theory,
that by proper modification of the off-diagonal elements of the
Kohn-Sham matrix these localizations can even be avoided in self-consistent
calculations for first-principles molecular
dynamics simulations \cite{Grimm2003,Friedrichs2008}.
The parameters used in these algorithms are case dependent
(localized / delocalized situations);
up to now no unique approach exists that would efficiently
converge to the correct solution for all cases.
Finally, if one wants to use Kohn-Sham instead of Hartree-Fock
expressions, there is the question how the energy expression for
a multi-determinant approach should be determined.
In ROKS \cite{Frank98} we use the energy expression as proposed in the
sum method \cite{Ziegler77}. This expression reduces to the ROHF terms if
it is used with the exact energy expectation values of the open-shell
Slater determinants, instead of inserting the Kohn-Sham expressions. \\

Practical ROKS dynamics calculations are presently
restricted to the case of two unpaired electrons which
is by far the most important case for the description of photoreactions.
This restriction has several reasons: 1. While the concept is clear,
there is, to our knowledge, no explicit formulation of the energy
expression for arbitrary low spin states in the Kohn-Sham literature.
2. Based on an energy expression, single-electron ROKS equations have to
be derived for the particular spin densities involved.
3. Algorithms for the self-consistent solution of the equations for N electrons must
be developed and tested. In the present paper, we address points 1 and 2. \\

We simplify the derivation, which corresponds to deriving an energy
expression for a spin-adapted configuration,
by neglecting the possibility of symmetry-determined
degeneracy as it may be present in atoms
($[\hat{H},\hat{L^2}]=0$) or in diatomic molecules ($[\hat{H},\hat{L_z}]=0$).
The treatment of such highly symmetric systems hardly plays a role
in molecular dynamics simulations. Instead we use the occupation pattern
as a symmetry like suggested in a similar way by the work of Ziegler,
Rauk and Baerends \cite{Ziegler77}, Daul \cite{Daul94}, and Noodleman \cite{Noodleman81,Noodleman86}.
On this basis it is possible to compute, for example not only the lowest singlet state,
but the singlet states with zero, two, and so on, unpaired electrons. (Note
that the singlet state with zero unpaired electrons is not necessarily the
lowest one.)
In calculations based on the sum method, in addition to the occupation pattern
high spatial symmetry was exploited for the computation of multiplets 
in the context of ligand field theory (see \cite{Atanasov2003}
and references cited therein).
We do not make use of radial or cylindrical symmetry (or
of approximate radial symmetry, if a central metal atom is in an environment).
This means that for example,
for two unpaired p electrons we cannot compute 
$^3$D, $^1$D, $^3$P, $^1P$, $^3S$, and $^1S$, but 
just T$_1$ and S$_1$. Higher states are only accessible as local minima.
Ignoring the fact that an atom might be in an environment with a particular
symmetry seems appropriate for reactive molecular dynamics simulations, where
the spatial symmetry and the degree of degeneracy should be free to change at any time.
Note that the restriction of not treating symmetry-determined degeneracy
explicitly, does not mean that degeneracy cannot be described at all.
It just means that degeneracy is determined at the single-configuration
level - which may turn out to be wrong in comparison with a multi-configuration approach. \\

In the next section, the energy expression for N unpaired electrons is
derived. In the following two sections, first the general form of the ROKS operators
for N electrons is given and then the explicit symmetry determined
parameters of these operators are specified for up to five unpaired electrons.

\section{Energy expression for N electrons}
\begin{figure}[H]
\centering
\includegraphics[height=11.5cm]{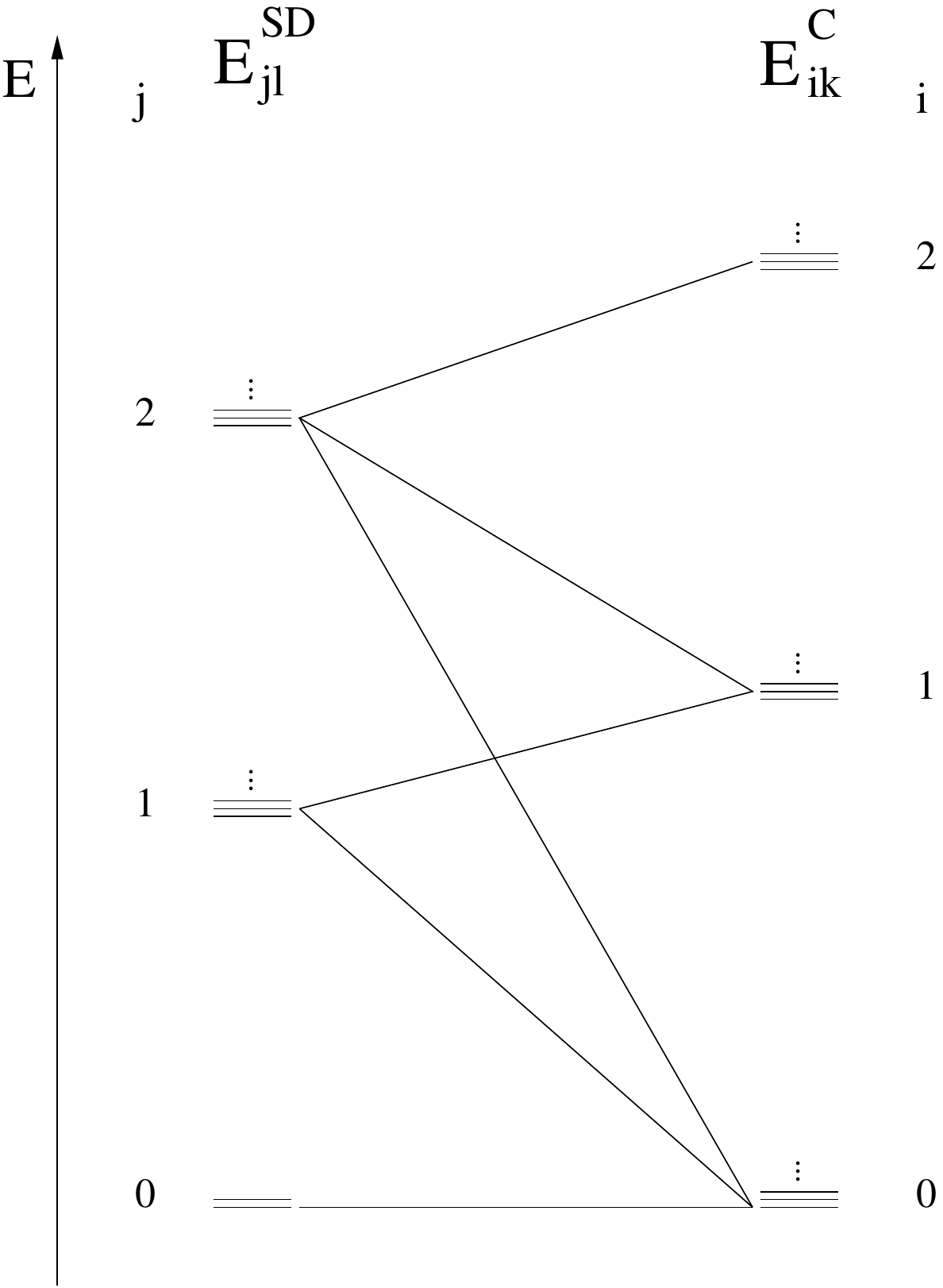}\\
\caption{\label{sketch}\small{Sketch of the relations between single-determinant
levels and single-configuration levels for an N electron case. 
The energy levels for a certain
multiplet $i$ (right side) are degenerate, the single-determinant energies are
averaged for a certain $j$ (corresponding to a certain
number of $\alpha$/$\beta$ electrons, left side). 
Hence the indices $l$ and $k$ numbering the different permutations / micro states
can be omitted
in the derivation of the general energy expression.
}}
\end{figure}

Since $[\hat{H},\hat{S}^2]=0$, an open-shell electronic configuration $\Psi_i^C$
must in general be composed of several Slater determinants $\Psi_j^{SD}$:

\begin{equation}
\label{eq:wave}
\Psi_i^C = \sum_j d_{ij} \Psi_j^{SD}
\end{equation}

In this way it is possible to describe a spin-pure energy state. 
For the derivation of the energy expression, the coefficients
$d_{ij}$ do not have to be explicitly determined. \\
The energy expectation values of a configuration $\Psi_i^C$ with N unpaired electrons
can be obtained as a sum of the energy expectation values of the single determinants:

\begin{equation}
 E_i^{C, N} = \sum_jc_{ij}E_j^{SD, N}
\end{equation}

In the derivation of the coefficients $c_{ij}$ we use the short notation:

\begin{equation}
 E^{C} = \sum_jc_jE_j^{SD}
\end{equation}

with

\begin{equation}
j = \frac{N - |n(\alpha)-n(\beta)|}{2}
\label{indexj}
\end{equation}

where $n(\alpha)$ and $n(\beta)$ are the total numbers of $\alpha$
and $\beta$ electrons respectively. \\
The Slater determinants which can be formed for a particular number 
of unpaired electrons N, can be ordered according to their
magnetic spin quantum number $M_S$:

\begin{equation}
M_S = \frac{1}{2}(n(\alpha)-n(\beta))
\end{equation}

\begin{figure}[H]
\centering
\includegraphics[height=6cm]{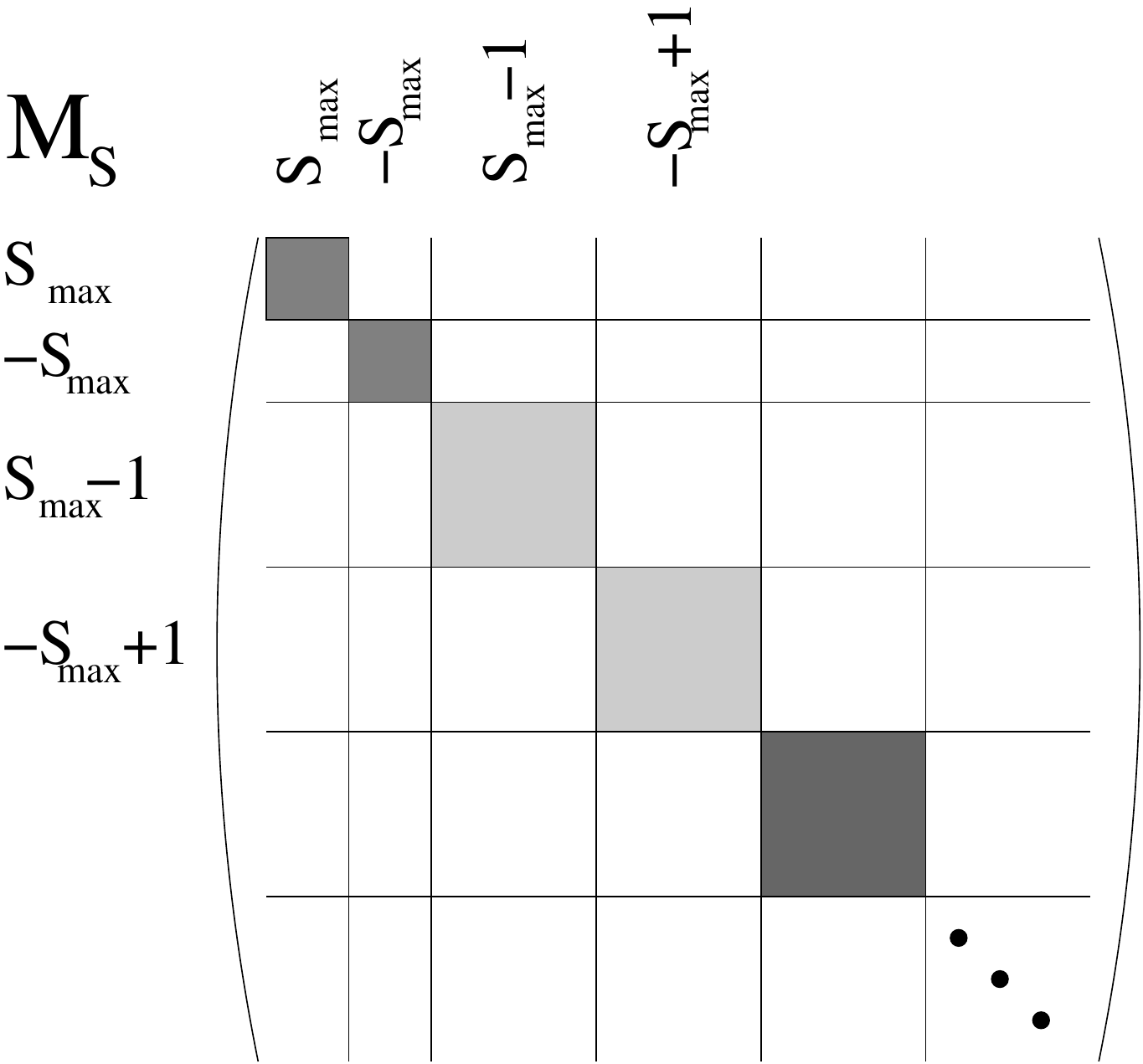}\\
\caption{\label{matrix}\small{
Hamilton matrix, ordered according to the $M_S$ values of the
Slater determinants. Only the elements in the colored blocks differ from zero.
}}
\end{figure}

With this ordering the Hamilton matrix for the
wavefunction \ref{eq:wave} factorizes (Fig. \ref{matrix}).

For S$_{max}$ and -S$_{max}$ (high-spin: all unpaired electrons
have either $\alpha$ or $\beta$ spin) the 1x1 block consists of the expression
for only one Slater determinant and $E^{C}$ = $E^{SD}_{j=0} = E^{SD}_{M_S=S_{max}}$ for the high-spin case.
From this, an energy expression can be obtained for the
block of the matrix with $M_S = S_{max}-1$:

\begin{small}
\begin{equation}
\label{eq:block1}
\binom{N}{1} \langle E_{M_S=S_{max}-1}^{SD} \rangle = \left[\binom{N}{1}-\binom{N}{0}\right]E^{S=S_{max}-1} + \binom{N}{0}E^{S=S_{max}}
\end{equation}
\end{small}

Similarly, for the block of the matrix with $M_S = S_{max}-2$:

\begin{scriptsize}
\begin{equation}
\binom{N}{2}\langle E_{M_S=S_{max}-2}^{SD} \rangle = \left[\binom{N}{2}-\binom{N}{1}\right]E^{S=S_{max}-2} + \left[\binom{N}{1}-\binom{N}{0}\right]E^{S=S_{max}-1} + \binom{N}{0}E^{S=S_{max}}
\end{equation}
\end{scriptsize}

With equation \ref{eq:block1} it follows:

\begin{small}
\begin{equation}
\binom{N}{2} \langle E_{M_S=S_{max}-2}^{SD} \rangle = \left[\binom{N}{2}-\binom{N}{1}\right]E^{S=S_{max}-2} + \binom{N}{1} \langle E_{M_S=S_{max}-1}^{SD}\rangle
\end{equation}
\end{small}

This result can be generalized for the matrix block with $M_S = S_{max}-j$:

\begin{small}
\begin{equation}
\binom{N}{j}\langle E_{M_S=S_{max}-j}^{SD} \rangle = \left[\binom{N}{j}-\binom{N}{j-1}\right]E^{S=S_{max}-j} + \binom{N}{j-1} \langle E_{M_S=S_{max}-(j-1)}^{SD}\rangle
\end{equation}
\end{small}

or:

\begin{small}
\begin{equation}
E^{S=S_{max}-j} =\frac{\binom{N}{j}\langle E_{M_S=S_{max}-j}^{SD}\rangle-\binom{N}{j-1}\langle E_{M_S=S_{max}-(j-1)}^{SD}\rangle }{\binom{N}{j}-\binom{N}{j-1}}
\end{equation}
\end{small}

In a shorter notation, we write:

\begin{small}
\begin{equation}
E^C_j =\frac{\binom{N}{j}\langle E_{j}^{SD}\rangle-\binom{N}{j-1}\langle E_{j-1}^{SD}\rangle }{\binom{N}{j}-\binom{N}{j-1}}
\end{equation}
\end{small}

From this formula it follows:
\begin{small}
\begin{equation}
E^C_j = \frac{N+1-j}{N+1-2j}\langle E_{j}^{SD}\rangle - \frac{j}{N+1-2j}\langle E_{j-1}^{SD}\rangle
\label{energyexpression}
\end{equation}
\end{small}

\begin{figure}[H]
\centering
\includegraphics[height=9cm]{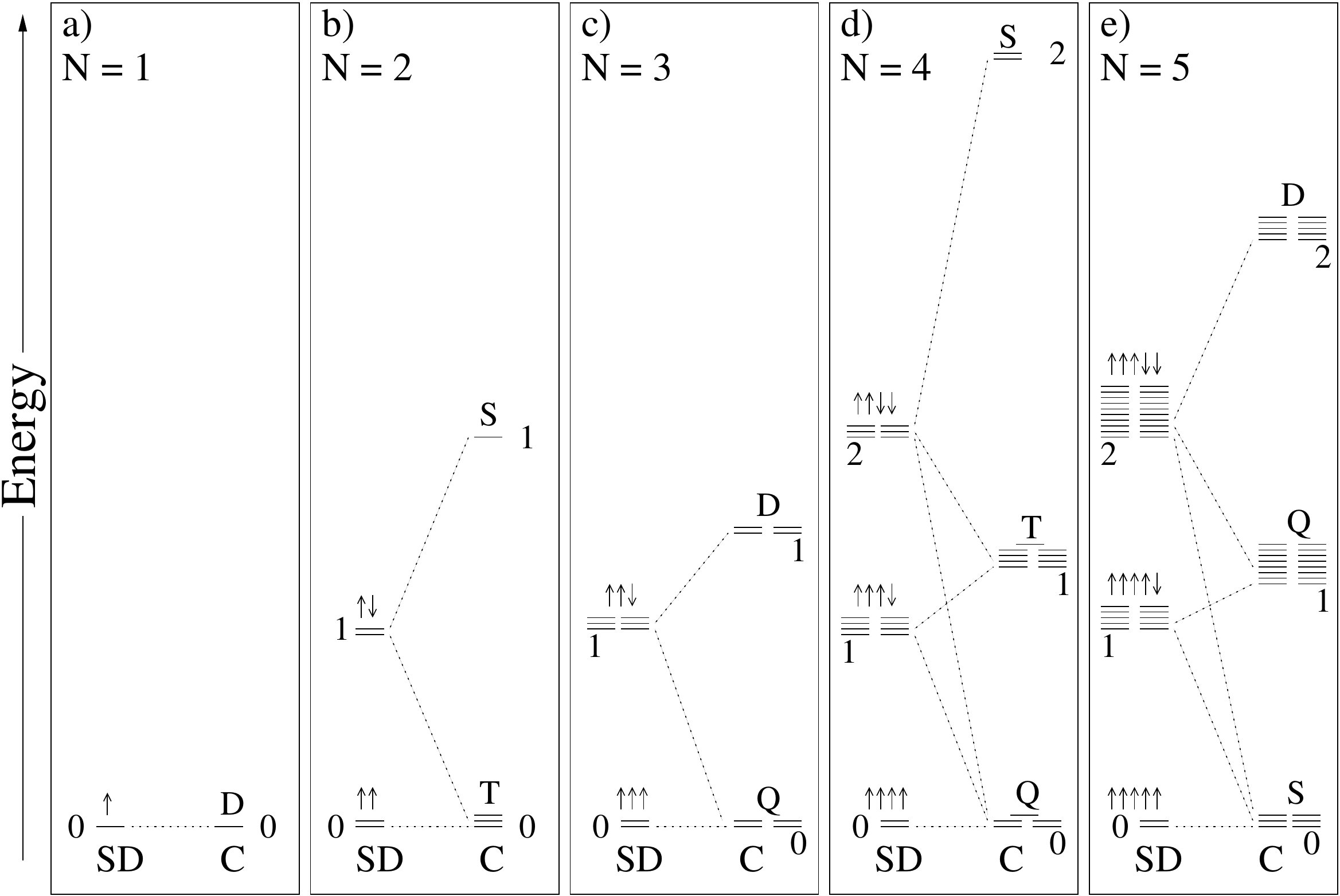}\\
\caption{\label{energydiagram}\small{Energy diagram for up to five unpaired electrons.
In the determination of the energy levels, equation \protect{\ref{energyexpression}}
was used. The energy levels are numbered by the index j.
}}
\end{figure}

That is, the energy of a state with multiplicity M = 2j + 1 can be determined
from the energies $E_{j}^{SD}$ and $E_{j-1}^{SD}$, all other coefficients
c$_j$ are zero. \\
The resulting energy levels are depicted true to scale in Fig. \ref{energydiagram}.

%%%%%%%%%%%%%%%%%%%%%%%%%%%%%%%%%%%%%%%%%%%%%%%%%5%%%%%%%%%%
For the computation of multiplet energies, Noodlemans elegant
and simple formula is frequently used \cite{Noodleman81}:

\begin{equation}
\label{eq:noodleman_energie}
 E^{LS} = (1+c) E^{BS} -cE^{HS}
\end{equation}

with

\begin{equation}
c = \frac{1}{S_{max} + S_{min}}
\end{equation}

In our notation this equation reads:

\begin{equation}
E^C_j = (1+c) E^{SD}_j -cE^{SD}_0
\end{equation}

In contrast to equation \ref{energyexpression}, 
the energies are determined from $E^{SD}_j$ and $E^{SD}_0$.
The Noodleman formula results in identical energy terms for up to three
unpaired electrons. (Two unpaired electrons: triplet: $E_0^C = E_0^{SD}$,
singlet: $E_1^C = 2E_1^{SD}-E_0^{SD}$, three unpaired electrons:
quartet: $E_0^C = E_0^{SD}$, 
doublet: $E_1^C = \frac{3}{2} E_1^{SD} - \frac{1}{2}E_0^{SD}$).
For more than three unpaired electrons, as is readily verified,
the Noodleman formula does not
fulfill the sum rule which says
that the sum of the energies of the configurations
must equal the sum of the energies of the determinants they are formed from.

\section{General ROKS equations}

Starting from

\begin{equation}
 E_j^C = c_{j}E_{j}+c_{j-1}E_{j-1}
\end{equation}

we derive single-electron ROKS equations.
Using the coefficients c$_j$ as determined by equation \ref{energyexpression},
the energy can be written in terms of energy expressions
of single Slater determinants,
whereby summation over half the determinants is sufficient since there are
always two determinants with equal energy for which the $\alpha$ and $\beta$
spins are just interchanged.

\begin{equation}
\begin{split}
 E_j^C = &c_{j}\frac{\sum_{l=1}^{\frac{a_{j}^{SD}}{2}}E_{jl}[\rho_{jl}^{\alpha},\rho_{jl}^{\beta}]}{\frac{a_j^{SD}}{2}}\\
&+c_{j-1}\frac{\sum_{l=1}^{\frac{a_{j-1}^{SD}}{2}}E_{(j-1)l}[\rho_{(j-1)l}^{\alpha},\rho_{(j-1)l}^{\beta}]}{\frac{a_{j-1}^{SD}}{2}}
\end{split}
\end{equation}

Using the Kohn-Sham energy expression

\begin{equation}
\label{summe2}
\begin{split}
E_j^C = &c_j \frac{\sum_{l=1}^{\frac{a_j^{SD}}{2}}\left[ T[\rho_{jl}] + J[\rho_{jl}] + E_{xc}[\rho_{jl}^{\alpha},\rho_{jl}^{\beta}] + \int v(\mathbf{r})\rho_{jl}(\mathbf{r}) d\mathbf{r} \right]}{\frac{a_j^{SD}}{2}}\\
&+c_{j-1} \frac{\sum_{l=1}^{\frac{a_{j-1}^{SD}}{2}}\left[ T[\rho_{(j-1)l}] + J[\rho_{(j-1)l}] + E_{xc}[\rho_{(j-1)l}^{\alpha},\rho_{(j-1)l}^{\beta}] + \int v(\mathbf{r})\rho_{(j-1)l}(\mathbf{r}) d\mathbf{r} \right]}{\frac{a_{j-1}^{SD}}{2}}
\end{split}
\end{equation}

the general ROKS operators for shell $i$ are obtained by functional variation:

\begin{scriptsize}
\begin{equation}
\begin{split}
\hat{F}_i = & \frac{1}{2}n_i\left[-\frac{1}{2}\nabla^2 + \int \frac{\rho(\mathbf{r'})}{|\mathbf{r}-\mathbf{r'}|} d\mathbf{r'} + v(\mathbf{r}) \right] + \frac{1}{2}c_j \frac{\sum_{l=1}^{\frac{a_j^{SD}}{2}}\left(n_{ijl}^{\alpha} v_{xc}^{\alpha}[\rho_{jl}^{\alpha},\rho_{jl}^{\beta}] + n_{ijl}^{\beta} v_{xc}^{\beta}[\rho_{jl}^{\alpha},\rho_{jl}^{\beta}]\right)}{\frac{a_j^{SD}}{2}}\\
&+\frac{1}{2}c_{j-1} \frac{\sum_{l=1}^{\frac{a_{j-1}^{SD}}{2}}\left(n_{i(j-1)l}^{\alpha} v_{xc}^{\alpha}[\rho_{(j-1)l}^{\alpha},\rho_{(j-1)l}^{\beta}] + n_{i(j-1)l}^{\beta} v_{xc}^{\beta}[\rho_{(j-1)l}^{\alpha},\rho_{(j-1)l}^{\beta}]\right)}{\frac{a_{j-1}^{SD}}{2}} 
\end{split}
\label{general}
\end{equation}
\end{scriptsize}

More details of the derivation are given for the case
of two unpaired electrons in \cite{Frank98}. \\
For practical applications of equation \ref{general}, the n$_{ijl}$ 
have to be determined, whereby the index $i$ denotes the - closed and
open - shells, the index $j$ which numbers the energy levels
is defined in equation \ref{indexj}, the index $l$ numbers
the microstates belonging to a certain $j$. The
coefficients n$_{ijl}$ are given in the following section
for up to five unpaired electrons.

%%%%%%%%%%%%%%%%%%%%%%%%%%%%%%%%%%%%%%%%%%%%%%%%%%%%%%%%%%%%%%%%%%%%%%%%%%555
\section{Special ROKS operators}
%%%%%%%%%%%%%%%%%%%%%%%%%%%%%%%%%%%%55

\subsection{Two unpaired electrons}

The case of two unpaired electrons is described in detail in \cite{Frank98}.
The energy expressions are as follows:

\begin{equation}
E^C_0 = E^{SD}_0
\end{equation}
\begin{equation}
E^C_1 = 2E^{SD}_1 - E^{SD}_0
\end{equation}

\begin{figure}[H]
\centering
\includegraphics[height=5cm]{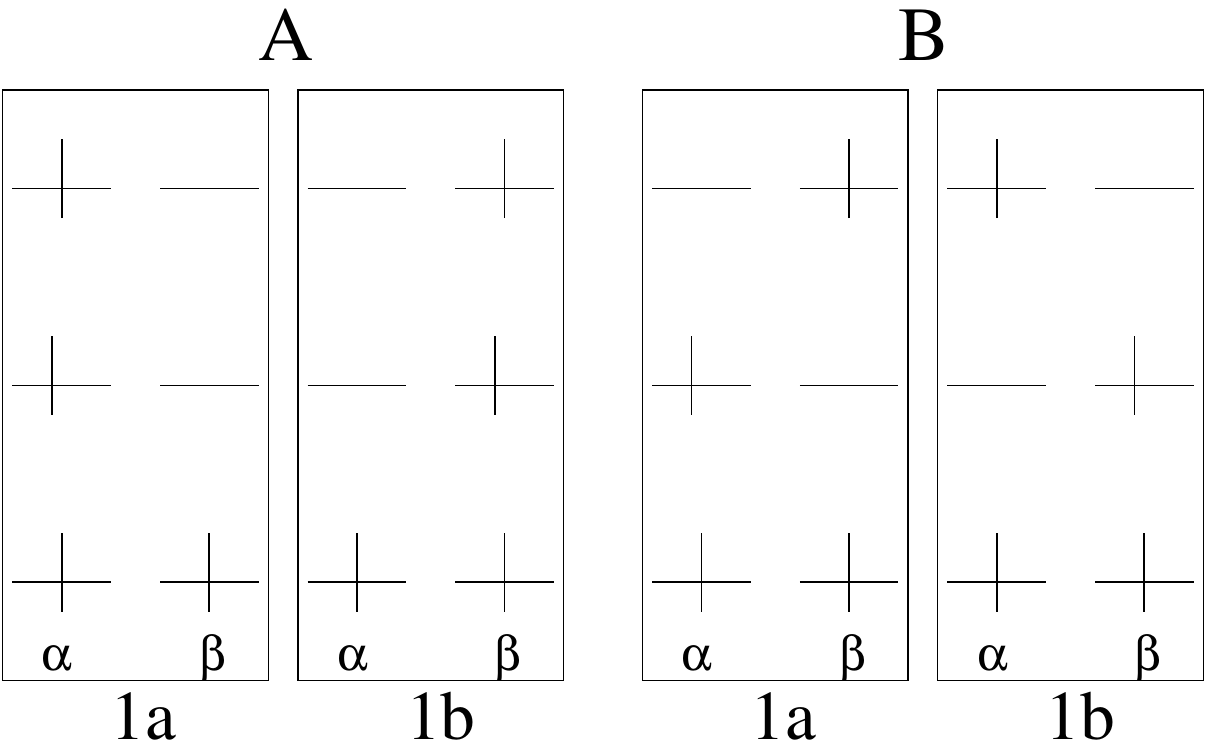}\\
\caption{\label{dens2}\small{Spin densities for the determinants that
can be formed for two unpaired electrons, grouped according to their
value of j. Here and in the following figures,
the indices l are described with 1a, 1b, 2a, 2b etc., indicating
that there are always two determinants with the same energy.
}}
\end{figure}

The spin densities are given in Fig. \ref{dens2}. The prefactors in
equation \ref{general} are listed in table \ref{tabspin2} for the non-trivial
singlet case.

%%%%
\begin{table}[H]
\centering
\begin{tabular}{|r||c|c||c|c|}
\hline
\multicolumn{1}{|c||}{} & \multicolumn{2}{|c||}{A} & \multicolumn{2}{|c|}{B} \\
No. & 1a & 1b & 1a & 1b\\
\hline
$n_c$ & 2 & 2 & 2 & 2\\
\hline
$n_{o1}^{\alpha}$ & 1 & 0 & 1 & 0\\
$n_{o1}^{\beta}$ & 0 & 1 & 0 & 1\\
\hline
$n_{o2}^{\alpha}$ & 1&0&0&1 \\
$n_{o2}^{\beta}$ & 0&1&1&0 \\
\hline
\end{tabular}
\caption{Coefficients in equation \protect{\ref{general}} for two unpaired electrons.}
\label{tabspin2}
\end{table}

This describes Kohn-Sham operators which explicitly read:

Triplet case:
\begin{equation}
\begin{split}
\hat{F}_c =& -\frac{1}{2}\nabla^2 + \int \frac{\rho(\mathbf{r'})}{|\mathbf{r}-\mathbf{r'}|} d\mathbf{r'} + v(\mathbf{r}) 
	+  \frac{1}{2}v_{xc}^{\alpha}[\rho_{A1}^{\alpha},\rho_{A1}^{\beta}] + \frac{1}{2}v_{xc}^{\beta}[\rho_{A1}^{\alpha},\rho_{A1}^{\beta}]\\
\hat{F}_{o1} =&  \frac{1}{2}\left[-\frac{1}{2}\nabla^2 + \int \frac{\rho(\mathbf{r'})}{|\mathbf{r}-\mathbf{r'}|} d\mathbf{r'} + v(\mathbf{r}) \right]
	+  \frac{1}{2}v_{xc}^{\alpha}[\rho_{A1}^{\alpha},\rho_{A1}^{\beta}]\\
\hat{F}_{o2} =&  \frac{1}{2}\left[-\frac{1}{2}\nabla^2 + \int \frac{\rho(\mathbf{r'})}{|\mathbf{r}-\mathbf{r'}|} d\mathbf{r'} + v(\mathbf{r}) \right]
	+  \frac{1}{2}v_{xc}^{\alpha}[\rho_{A1}^{\alpha},\rho_{A1}^{\beta}]\\
\end{split}
\end{equation}

Singlet case:
\begin{equation}
\begin{split}
\hat{F}_c =& -\frac{1}{2}\nabla^2 + \int \frac{\rho(\mathbf{r'})}{|\mathbf{r}-\mathbf{r'}|} d\mathbf{r'} + v(\mathbf{r})
	+ v_{xc}^{\alpha}[\rho_{B1}^{\alpha},\rho_{B1}^{\beta}] + v_{xc}^{\beta}[\rho_{B1}^{\alpha},\rho_{B1}^{\beta}]\\
	&- \frac{1}{2} (v_{xc}^{\alpha}[\rho_{A1}^{\alpha},\rho_{A1}^{\beta}] + v_{xc}^{\beta}[\rho_{A1}^{\alpha},\rho_{A1}^{\beta}])\\
\hat{F}_{o1} =&  \frac{1}{2} \left[-\frac{1}{2}\nabla^2 + \int \frac{\rho(\mathbf{r'})}{|\mathbf{r}-\mathbf{r'}|} d\mathbf{r'} + v(\mathbf{r}) \right]
	+  v_{xc}^{\alpha}[\rho_{B1}^{\alpha},\rho_{B1}^{\beta}]
	-  \frac{1}{2} v_{xc}^{\alpha}[\rho_{A1}^{\alpha},\rho_{A1}^{\beta}]\\
\hat{F}_{o2} =&  \frac{1}{2} \left[-\frac{1}{2}\nabla^2 + \int \frac{\rho(\mathbf{r'})}{|\mathbf{r}-\mathbf{r'}|} d\mathbf{r'} + v(\mathbf{r}) \right]
	+  v_{xc}^{\beta}[\rho_{B1}^{\alpha},\rho_{B1}^{\beta}]
	-  \frac{1}{2} v_{xc}^{\alpha}[\rho_{A1}^{\alpha},\rho_{A1}^{\beta}]\\
\end{split}
\end{equation}

\subsection{Three unpaired electrons}

\begin{equation}
E^C_0 = E^{SD}_0
\end{equation}
\begin{equation}
E^C_1 = \frac{3}{2}E^{SD}_1 - \frac{1}{2}E^{SD}_0
\end{equation}

\begin{figure}[H]
\centering
\includegraphics[height=5cm]{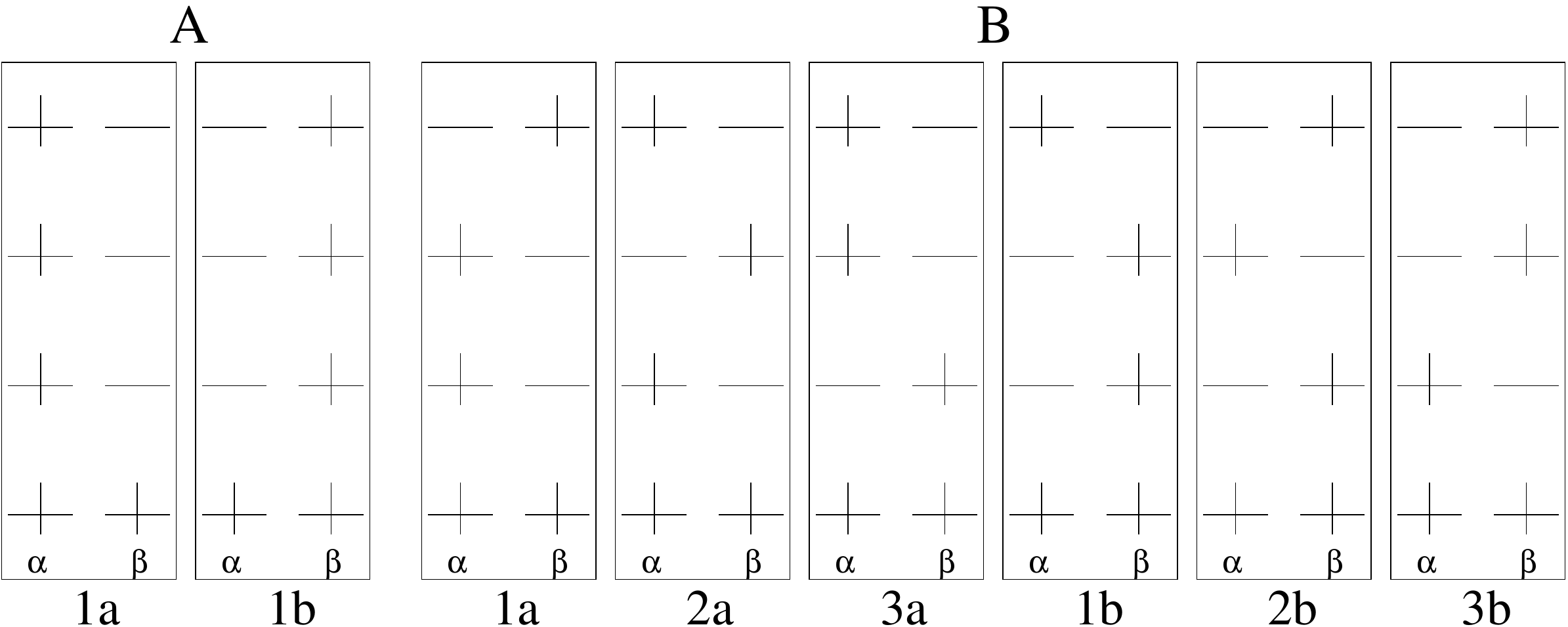}\\
\caption{\label{dens3}\small{Spin densities for three unpaired electrons.}}
\end{figure}
%%%

\begin{table}[H]
\centering
\begin{tabular}{|r||c|c||c|c|c|c|c|c|}
\hline
\multicolumn{1}{|c||}{} & \multicolumn{2}{|c||}{A} & \multicolumn{6}{|c|}{B} \\
No. & 1a & 1b & 1a & 2a & 3a & 1b & 2b & 3b\\
\hline
$n_c$ & 2 & 2 & 2 & 2&2&2&2&2\\
\hline
$n_{o1}^{\alpha}$ & 1&0&1&1&0&0&0&1\\
$n_{o1}^{\beta}$ &  0&1&0&0&1&1&1&0\\
\hline
$n_{o2}^{\alpha}$ & 1&0&1&0&1&0&1&0\\
$n_{o2}^{\beta}$ &  0&1&0&1&0&1&0&1\\
\hline
$n_{o3}^{\alpha}$ & 1&0&0&1&1&1&0&0\\
$n_{o3}^{\beta}$ &  0&1&1&0&0&0&1&1\\
\hline
\end{tabular}
\caption{Coefficients in equation \protect{\ref{general}} for three unpaired electrons.}
\label{tabspin3}
\end{table}

The Kohn-Sham operators for three unpaired electrons (quartet and doublet) are:

Quartet case:
\begin{equation}
\begin{split}
\hat{F}_c =& -\frac{1}{2}\nabla^2 + \int \frac{\rho(\mathbf{r'})}{|\mathbf{r}-\mathbf{r'}|} d\mathbf{r'} + v(\mathbf{r}) 
	+ \frac{1}{2} v_{xc}^{\alpha}[\rho_{A1}^{\alpha},\rho_{A1}^{\beta}] + \frac{1}{2} v_{xc}^{\beta}[\rho_{A1}^{\alpha},\rho_{A1}^{\beta}] \\
\hat{F}_{o1} =&  \frac{1}{2} \left[-\frac{1}{2}\nabla^2 + \int \frac{\rho(\mathbf{r'})}{|\mathbf{r}-\mathbf{r'}|} d\mathbf{r'} + v(\mathbf{r}) \right]
	+  \frac{1}{2} v_{xc}^{\alpha}[\rho_{A1}^{\alpha},\rho_{A1}^{\beta}]\\
\hat{F}_{o2} =&  \frac{1}{2} \left[-\frac{1}{2}\nabla^2 + \int \frac{\rho(\mathbf{r'})}{|\mathbf{r}-\mathbf{r'}|} d\mathbf{r'} + v(\mathbf{r}) \right]
	+  \frac{1}{2} v_{xc}^{\alpha}[\rho_{A1}^{\alpha},\rho_{A1}^{\beta}]\\
\hat{F}_{o3} =&  \frac{1}{2} \left[-\frac{1}{2}\nabla^2 + \int \frac{\rho(\mathbf{r'})}{|\mathbf{r}-\mathbf{r'}|} d\mathbf{r'} + v(\mathbf{r}) \right]
	+ \frac{1}{2} v_{xc}^{\alpha}[\rho_{A1}^{\alpha},\rho_{A1}^{\beta}]\\
\end{split}
\end{equation}

Doublet case:
\begin{equation}
\begin{split}
\hat{F}_c =& -\frac{1}{2}\nabla^2 + \int \frac{\rho(\mathbf{r'})}{|\mathbf{r}-\mathbf{r'}|} d\mathbf{r'} + v(\mathbf{r}) \\
	&+ \frac{3}{4} \Bigl( \frac{1}{3} (v_{xc}^{\alpha}[\rho_{B1}^{\alpha},\rho_{B1}^{\beta}] + v_{xc}^{\beta}[\rho_{B1}^{\alpha},\rho_{B1}^{\beta}]
			+ v_{xc}^{\alpha}[\rho_{B2}^{\alpha},\rho_{B2}^{\beta}]\\
&~~~~~~~~~~+ v_{xc}^{\beta}[\rho_{B2}^{\alpha},\rho_{B2}^{\beta}] 
			+ v_{xc}^{\alpha}[\rho_{B3}^{\alpha},\rho_{B3}^{\beta}] + v_{xc}^{\beta}[\rho_{B3}^{\alpha},\rho_{B3}^{\beta}]) \Bigr) \\
	&- \frac{1}{4} (v_{xc}^{\alpha}[\rho_{A1}^{\alpha},\rho_{A1}^{\beta}] + v_{xc}^{\beta}[\rho_{A1}^{\alpha},\rho_{A1}^{\beta}]) \\
\hat{F}_{o1} =& \frac{1}{2} \left[-\frac{1}{2}\nabla^2 + \int \frac{\rho(\mathbf{r'})}{|\mathbf{r}-\mathbf{r'}|} d\mathbf{r'} + v(\mathbf{r}) \right]\\
	&+ \frac{3}{4}\left( \frac{1}{3} (v_{xc}^{\alpha}[\rho_{B1}^{\alpha},\rho_{B1}^{\beta}] 
			+ v_{xc}^{\alpha}[\rho_{B2}^{\alpha},\rho_{B2}^{\beta}] 
			+ v_{xc}^{\beta}[\rho_{B3}^{\alpha},\rho_{B3}^{\beta}])\right)\\
	&- \frac{1}{4} v_{xc}^{\alpha}[\rho_{A1}^{\alpha},\rho_{A1}^{\beta}]\\
\hat{F}_{o2} =& \frac{1}{2} \left[-\frac{1}{2}\nabla^2 + \int \frac{\rho(\mathbf{r'})}{|\mathbf{r}-\mathbf{r'}|} d\mathbf{r'} + v(\mathbf{r}) \right]\\
	&+ \frac{3}{4}\left( \frac{1}{3} (v_{xc}^{\alpha}[\rho_{B1}^{\alpha},\rho_{B1}^{\beta}]
			 + v_{xc}^{\beta}[\rho_{B2}^{\alpha},\rho_{B2}^{\beta}]
			+ v_{xc}^{\alpha}[\rho_{B3}^{\alpha},\rho_{B3}^{\beta}])\right)\\
	&- \frac{1}{4} v_{xc}^{\alpha}[\rho_{A1}^{\alpha},\rho_{A1}^{\beta}]\\
\hat{F}_{o3} =& \frac{1}{2} \left[-\frac{1}{2}\nabla^2 + \int \frac{\rho(\mathbf{r'})}{|\mathbf{r}-\mathbf{r'}|} d\mathbf{r'} + v(\mathbf{r}) \right]\\
	&+ \frac{3}{4}\left( \frac{1}{3} (v_{xc}^{\beta}[\rho_{B1}^{\alpha},\rho_{B1}^{\beta}] 
			+ v_{xc}^{\alpha}[\rho_{B2}^{\alpha},\rho_{B2}^{\beta}]
			+ v_{xc}^{\alpha}[\rho_{B3}^{\alpha},\rho_{B3}^{\beta}])\right)\\
	&- \frac{1}{4} v_{xc}^{\alpha}[\rho_{A1}^{\alpha},\rho_{A1}^{\beta}]\\
\end{split}
\end{equation}

To simplify the representation for the cases with more electrons we introduce a
short notation for these equations (Tables \ref{taboperator3quartet} and \ref{taboperator3doublet}).

\begin{table}[H]
\begin{scriptsize}
\centering
\begin{tabular}{|r||c|c|c|c|c|}
\hline
& \multicolumn{2}{|c|}{} & \multicolumn{2}{|c|}{A} &\\
& \multicolumn{2}{|c|}{} & \multicolumn{2}{|c|}{1}  &\\
 & $\frac{1}{2}[~]$ & $+\frac{1}{2}[($ & $\alpha$ & $\beta$ & $)]$\\
\hline
c  & 2 &&  1&1&\\
o1 & 1 &&  1&0&\\
o2 & 1 &&  1&0&\\
o3 & 1 &&  1&0&\\
\hline
\end{tabular}
\caption{ROKS operators for three unpaired electrons, quartet case.}
\label{taboperator3quartet}
\end{scriptsize}
\end{table}

\begin{table}[H]
\begin{scriptsize}
\centering
\begin{tabular}{|r||c|c|c|c|c|c|c|c|c|c|c|c|}
\hline
&\multicolumn{2}{|c|}{} & \multicolumn{6}{|c|}{B} & & \multicolumn{2}{|c|}{A} &\\
&\multicolumn{2}{|c|}{} & \multicolumn{2}{|c|}{1} & \multicolumn{2}{|c|}{2} & \multicolumn{2}{|c|}{3} & & \multicolumn{2}{|c|}{1}  &\\
 & $\frac{1}{2}[~]$ & $+\frac{3}{4}[\frac{1}{3}($ & $\alpha$ & $\beta$ & $\alpha$ & $\beta$ & $\alpha$ & $\beta$ & $)]-\frac{1}{4}[($ & $\alpha$ & $\beta$ & $)]$\\
\hline
c  & 2 & & 1&1&1&1&1&1 &&  1&1&\\
o1 & 1 & & 1&0&1&0&0&1 &&  1&0&\\
o2 & 1 & & 1&0&0&1&1&0 &&  1&0&\\
o3 & 1 & & 0&1&1&0&1&0 &&  1&0&\\
\hline
\end{tabular}
\caption{ROKS operators for three unpaired electrons, doublet case.}
\label{taboperator3doublet}
\end{scriptsize}
\end{table}

\subsection{Four unpaired electrons}

\begin{equation}
E^C_0 = E^{SD}_0
\end{equation}
\begin{equation}
E^C_1 = \frac{4}{3}E^{SD}_1 - \frac{1}{3}E^{SD}_0
\end{equation}
\begin{equation}
E^C_2 = 3E^{SD}_2 - 2E^{SD}_1
\end{equation}

\begin{figure}[H]
\centering
\includegraphics[height=10cm]{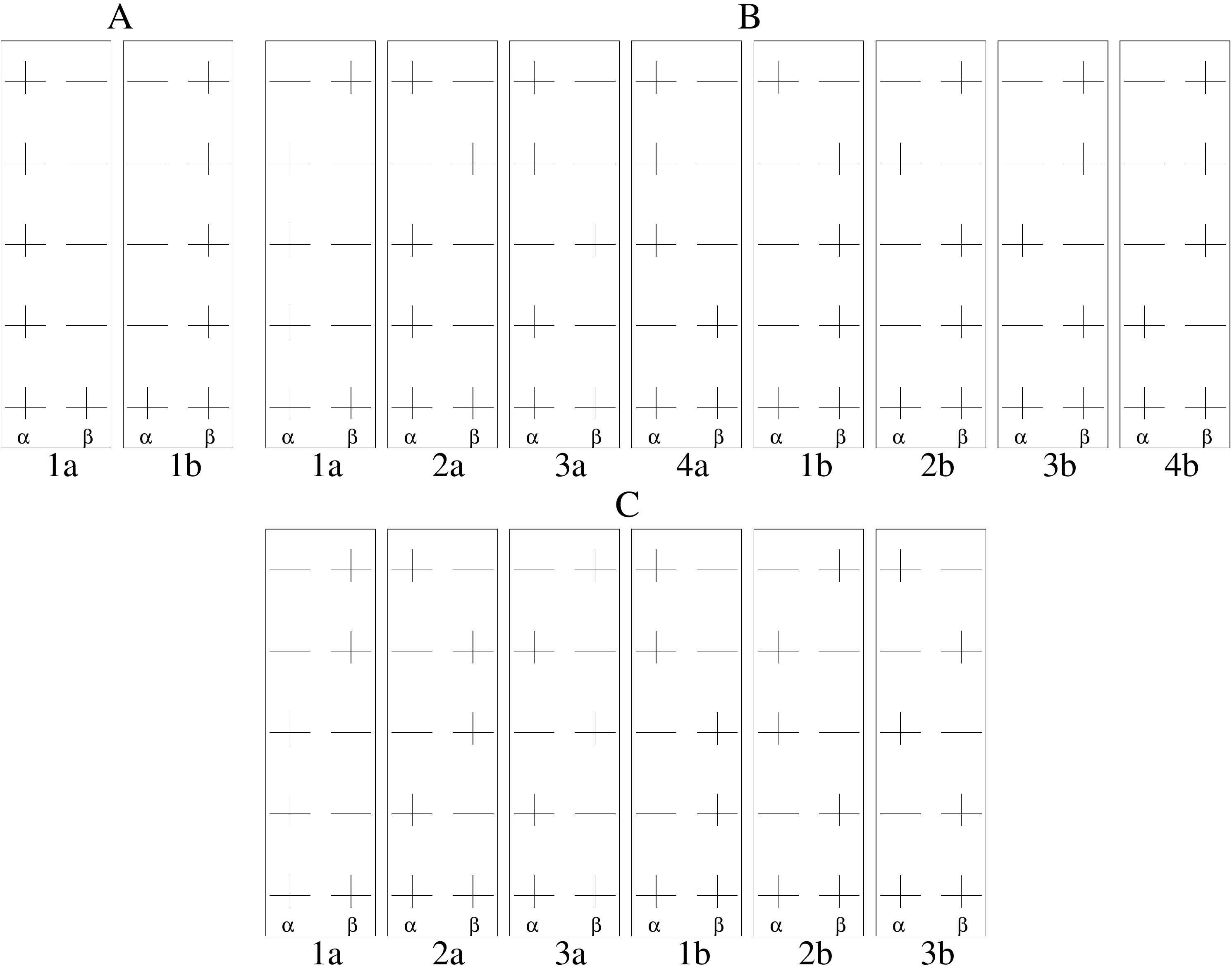}\\
\caption{\label{dens4}\small{Spin densities for four unpaired electrons.}}
\end{figure}
%%%

Table \ref{tabspin4} lists the coefficients for
four unpaired electrons, tables \ref{taboperator4quintet}, \ref{taboperator4triplet} and
\ref{taboperator4singlet} the Fock operators in short notation.

\begin{table}[H]
\centering
\begin{tabular}{|r||c|c||c|c|c|c|c|c|c|c||c|c|c|c|c|c|}
\hline
\multicolumn{1}{|c||}{} & \multicolumn{2}{|c||}{A} & \multicolumn{8}{|c||}{B} & \multicolumn{6}{|c|}{C}\\
No. & 1a & 1b & 1a & 2a & 3a & 4a & 1b & 2b & 3b & 4b & 1a & 2a & 3a & 1b & 2b & 3b \\
\hline
$n_c$ & 2&2&2&2&2&2&2&2&2&2&2&2&2&2&2&2\\
\hline
$n_{o1}^{\alpha}$ & 1&0&1&1&1&0&0&0&0&1&1&1&1&0&0&0\\
$n_{o1}^{\beta}$ &  0&1&0&0&0&1&1&1&1&0&0&0&0&1&1&1\\
\hline
$n_{o2}^{\alpha}$ & 1&0&1&1&0&1&0&0&1&0&1&0&0&0&1&1\\
$n_{o2}^{\beta}$ &  0&1&0&0&1&0&1&1&0&1&0&1&1&1&0&0\\
\hline
$n_{o3}^{\alpha}$ & 1&0&1&0&1&1&0&1&0&0&0&0&1&1&1&0\\
$n_{o3}^{\beta}$ &  0&1&0&1&0&0&1&0&1&1&1&1&0&0&0&1\\
\hline
$n_{o4}^{\alpha}$ & 1&0&0&1&1&1&1&0&0&0&0&1&0&1&0&1\\
$n_{o4}^{\beta}$ &  0&1&1&0&0&0&0&1&1&1&1&0&1&0&1&0\\
\hline
\end{tabular}
\caption{Coefficients in equation \protect{\ref{general}} for four unpaired electrons.}
\label{tabspin4}
\end{table}

\begin{table}[H]
\begin{scriptsize}
\centering
\begin{tabular}{|r||c|c|c|c|c|}
\hline
&\multicolumn{2}{|c|}{} &  \multicolumn{2}{|c|}{A} &\\
&\multicolumn{2}{|c|}{} &  \multicolumn{2}{|c|}{1}  &\\
 & $\frac{1}{2}[~]$ & $+\frac{1}{2}[($ & $\alpha$ & $\beta$ & $)]$\\
\hline
c  & 2 & &  1&1&\\
o1 & 1 & &  1&0&\\
o2 & 1 & &  1&0&\\
o3 & 1 & &  1&0&\\
o4 & 1 & &  1&0&\\
\hline
\end{tabular}
\caption{ROKS operators for four unpaired electrons, quintet case.}
\label{taboperator4quintet}
\end{scriptsize}
\end{table}
%%%%%%%%
\begin{table}[H]
\begin{scriptsize}
\centering
\begin{tabular}{|r||c|c|c|c|c|c|c|c|c|c|c|c|c|c|}
\hline
&\multicolumn{2}{|c|}{} & \multicolumn{8}{|c|}{B} & & \multicolumn{2}{|c|}{A} &\\
&\multicolumn{2}{|c|}{} & \multicolumn{2}{|c|}{1} & \multicolumn{2}{|c|}{2} & \multicolumn{2}{|c|}{3} & \multicolumn{2}{|c|}{4} & & \multicolumn{2}{|c|}{1}  &\\
 & $\frac{1}{2}[~]$ & $+\frac{4}{6}[\frac{1}{4}($ & $\alpha$ & $\beta$ & $\alpha$ & $\beta$ & $\alpha$ & $\beta$ & $\alpha$ & $\beta$ & $)]-\frac{1}{6}[($ & $\alpha$ & $\beta$ & $)]$\\
\hline
c  & 2 & & 1&1&1&1&1&1&1&1 &&  1&1&\\
o1 & 1 & & 1&0&1&0&1&0&0&1 &&  1&0&\\
o2 & 1 & & 1&0&1&0&0&1&1&0 &&  1&0&\\
o3 & 1 & & 1&0&0&1&1&0&1&0 &&  1&0&\\
o4 & 1 & & 0&1&1&0&1&0&1&0 &&  1&0&\\
\hline
\end{tabular}
\caption{ROKS operators for four unpaired electrons, triplet case.}
\label{taboperator4triplet}
\end{scriptsize}
\end{table}
%%%%%%%%
\begin{table}[H]
\begin{scriptsize}
\centering
\begin{tabular}{|r||c|c|c|c|c|c|c|c|c|c|c|c|c|c|c|c|c|c|}
\hline
&\multicolumn{2}{|c|}{} & \multicolumn{6}{|c|}{C} & & \multicolumn{8}{|c|}{B} &\\
&\multicolumn{2}{|c|}{} & \multicolumn{2}{|c|}{1} & \multicolumn{2}{|c|}{2} & \multicolumn{2}{|c|}{3}  & & \multicolumn{2}{|c|}{1}  & \multicolumn{2}{|c|}{2}& \multicolumn{2}{|c|}{3}& \multicolumn{2}{|c|}{4}&\\
 & $\frac{1}{2}[~]$ & $+\frac{3}{2}[\frac{1}{3}($ & $\alpha$ & $\beta$ & $\alpha$ & $\beta$ & $\alpha$ & $\beta$ & $)]-[\frac{1}{4}($& $\alpha$ & $\beta$& $\alpha$ & $\beta$& $\alpha$ & $\beta$ & $\alpha$ & $\beta$ & $)]$\\
\hline
c  & 2 && 1&1&1&1&1&1 && 1&1&1&1&1&1&1&1 &\\
o1 & 1 && 1&0&1&0&1&0 && 1&0&1&0&1&0&0&1 &\\
o2 & 1 && 1&0&0&1&0&1 && 1&0&1&0&0&1&1&0 &\\
o3 & 1 && 0&1&0&1&1&0 && 1&0&0&1&1&0&1&0 &\\
o4 & 1 && 0&1&1&0&0&1 && 0&1&1&0&1&0&1&0 &\\
\hline
\end{tabular}
\caption{ROKS operators for four unpaired electrons, singlet case.}
\label{taboperator4singlet}
\end{scriptsize}
\end{table}

\subsection{Five unpaired electrons}

\begin{equation}
E^C_0 = E^{SD}_0
\end{equation}
\begin{equation}
E^C_1 = \frac{5}{4}E^{SD}_1 - \frac{1}{4}E^{SD}_0
\end{equation}
\begin{equation}
E^C_2 = 2E^{SD}_2 - E^{SD}_1
\end{equation}

\begin{figure}[H]
\centering
\includegraphics[height=15cm]{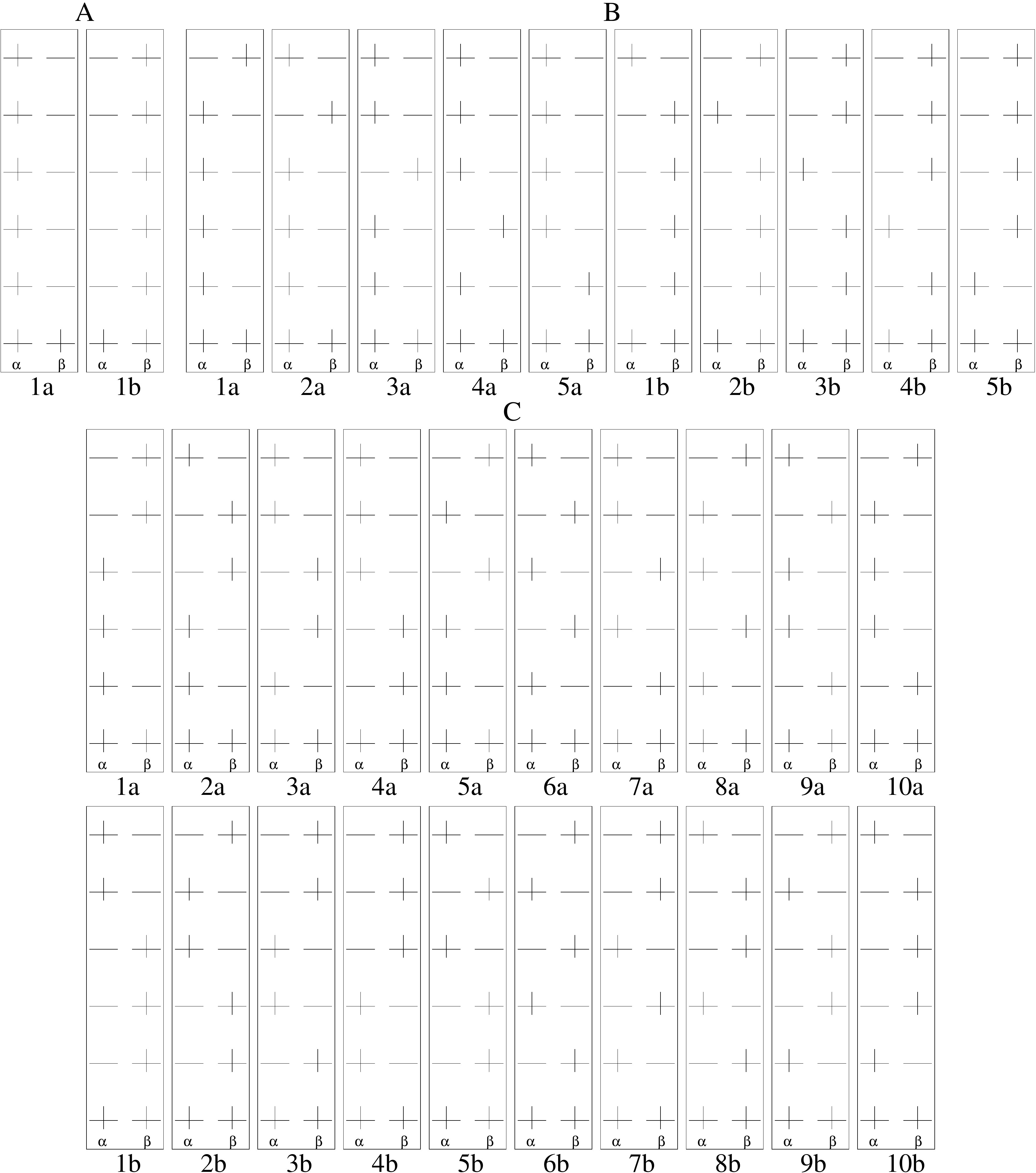}\\
\caption{\label{dens_5.eps}\small{Spin densities for five unpaired electrons.}}
\end{figure}
%%%

Table \ref{tabspin5} lists the coefficients for
five unpaired electrons, tables \ref{taboperator5sextet}, \ref{taboperator5quartet} and
\ref{taboperator5doublet} the Fock operators in short notation.

\begin{table}[H]
\centering
\begin{scriptsize}
\begin{tabular}{|r||c|c||c|c|c|c|c|c|c|c|c|c||c|c|c|c|c|c|c|c|c|c|c|c|c|c|c|c|c|c|c|c|c|c|c|c|}
\hline
\multicolumn{1}{|c||}{} & \multicolumn{2}{|c||}{A} & \multicolumn{10}{|c||}{B} & \multicolumn{20}{|c|}{C}\\
No. & 1a & 1b & 1a & 2a & 3a & 4a & 5a & 1b & 2b & 3b & 4b & 5b & 1a & 2a & 3a & 4a & 5a & 6a & 7a & 8a & 9a & 10a & 1b & 2b & 3b & 4b & 5b & 6b & 7b & 8b & 9b & 10b \\
\hline
$n_c$ & 2&2&2&2&2&2&2&2&2&2&2&2  & 2&2&2&2&2&2&2&2&2&2&2&2&2&2&2&2&2&2&2&2\\
\hline
$n_{o1}^{\alpha}$ & 1&0&1&1&1&1&0&0&0&0&0&1 & 1&1&1&0&1&1&0&1&0&0&0&0&0&1&0&0&1&0&1&1\\
$n_{o1}^{\beta}$ &  0&1&0&0&0&0&1&1&1&1&1&0 &  0&0&0&1&0&0&1&0&1&1&1&1&1&0&1&1&0&1&0&0\\
\hline
$n_{o2}^{\alpha}$ & 1&0&1&1&1&0&1&0&0&0&1&0 & 1&1&0&0&1&0&1&0&1&1&0&0&1&1&0&1&0&1&0&0\\
$n_{o2}^{\beta}$ &  0&1&0&0&0&1&0&1&1&1&0&1 &  0&0&1&1&0&1&0&1&0&0&1&1&0&0&1&0&1&0&1&1\\
\hline
$n_{o3}^{\alpha}$ & 1&0&1&1&0&1&1&0&0&1&0&0 & 1&0&0&1&0&1&0&1&1&1&0&1&1&0&1&0&1&0&0&0\\
$n_{o3}^{\beta}$ &  0&1&0&0&1&0&0&1&1&0&1&1 &  0&1&1&0&1&0&1&0&0&0&1&0&0&1&0&1&0&1&1&1\\
\hline
$n_{o4}^{\alpha}$ & 1&0&1&0&1&1&1&0&1&0&0&0 & 0&0&1&1&1&0&1&1&0&1&1&1&0&0&0&1&0&0&1&0\\
$n_{o4}^{\beta}$ &  0&1&0&1&0&0&0&1&0&1&1&1 &  1&1&0&0&0&1&0&0&1&0&0&0&1&1&1&0&1&1&0&1\\
\hline
$n_{o5}^{\alpha}$ & 1&0&0&1&1&1&1&1&0&0&0&0 & 0&1&1&1&0&1&1&0&1&0&1&0&0&0&1&0&0&1&0&1\\
$n_{o5}^{\beta}$ &  0&1&1&0&0&0&0&0&1&1&1&1 &  1&0&0&0&1&0&0&1&0&1&0&1&1&1&0&1&1&0&1&0\\
\hline
\end{tabular}
\end{scriptsize}
\caption{Coefficients in equation \protect{\ref{general}} for five unpaired electrons.}
\label{tabspin5}
\end{table}

\begin{table}[H]
\centering
\begin{scriptsize}
\begin{tabular}{|r||c|c|c|c|c|c|}
\hline
&\multicolumn{2}{|c|}{} & \multicolumn{2}{|c|}{A} &\\
&\multicolumn{2}{|c|}{} & \multicolumn{2}{|c|}{1} &\\
 & $\frac{1}{2}[~]$ & $+\frac{1}{2}[($ & $\alpha$ & $\beta$ & $)]$\\
\hline
c  & 2 & &  1&1&\\
o1 & 1 & &  1&0&\\
o2 & 1 & &  1&0&\\
o3 & 1 & &  1&0&\\
o4 & 1 & &  1&0&\\
o5 & 1 & &  1&0&\\
\hline
\end{tabular}
\caption{ROKS operators for five unpaired electrons, sextet case.}
\label{taboperator5sextet}
\end{scriptsize}
\end{table}
%%%%%%%%%%%%%%%%
\begin{table}[H]
\centering
\begin{scriptsize}
\begin{tabular}{|r||c|c|c|c|c|c|c|c|c|c|c|c|c|c|c|c|c|}
\hline
&\multicolumn{2}{|c|}{} & \multicolumn{10}{|c|}{B} & & \multicolumn{2}{|c|}{A} &\\
&\multicolumn{2}{|c|}{} & \multicolumn{2}{|c|}{1} & \multicolumn{2}{|c|}{2} & \multicolumn{2}{|c|}{3} & \multicolumn{2}{|c|}{4} & \multicolumn{2}{|c|}{5} && \multicolumn{2}{|c|}{1}  &\\
 & $\frac{1}{2}[~]$ & $+\frac{5}{8}[\frac{1}{5}($ & $\alpha$ & $\beta$ & $\alpha$ & $\beta$ & $\alpha$ & $\beta$ & $\alpha$ & $\beta$ & $\alpha$ & $\beta$ & $)]-\frac{1}{8}[($ & $\alpha$ & $\beta$ & $)]$\\
\hline
c  & 2 & & 1&1&1&1&1&1&1&1&1&1 &&  1&1&\\
o1 & 1 & & 1&0&1&0&1&0&1&0&0&1 &&  1&0&\\
o2 & 1 & & 1&0&1&0&1&0&0&1&1&0 &&  1&0&\\
o3 & 1 & & 1&0&1&0&0&1&1&0&1&0 &&  1&0&\\
o4 & 1 & & 1&0&0&1&1&0&1&0&1&0 &&  1&0&\\
o5 & 1 & & 0&1&1&0&1&0&1&0&1&0 &&  1&0&\\
\hline
\end{tabular}
\caption{ROKS operators for five unpaired electrons, quartet case.}
\label{taboperator5quartet}
\end{scriptsize}
\end{table}
%
%%%%%%%%%%%%%%%%
\begin{table}[H]
\begin{scriptsize}
\centering
%\begin{rotate}{90}
\begin{tabular}{|r||c|c|c|c|c|c|c|c|c|c|c|c|c|c|c|c|c|c|c|c|c|c|c|c|c|c|c|c|c|c|c|c|c|c|}
\hline
& \multicolumn{2}{|c|}{} & \multicolumn{20}{|c|}{C} &  & \multicolumn{10}{|c|}{B} &\\
& \multicolumn{2}{|c|}{} & \multicolumn{2}{|c|}{1} & \multicolumn{2}{|c|}{2} & \multicolumn{2}{|c|}{3} & \multicolumn{2}{|c|}{4} & \multicolumn{2}{|c|}{5} & \multicolumn{2}{|c|}{6} & \multicolumn{2}{|c|}{7} & \multicolumn{2}{|c|}{8} & \multicolumn{2}{|c|}{9} & \multicolumn{2}{|c|}{10} &  & \multicolumn{2}{|c|}{1} & \multicolumn{2}{|c|}{2} & \multicolumn{2}{|c|}{3} & \multicolumn{2}{|c|}{4} & \multicolumn{2}{|c|}{5} &\\
 & $\frac{1}{2}[~]$ & $+[\frac{1}{10}($ & $\alpha$ & $\beta$ & $\alpha$ & $\beta$ & $\alpha$ & $\beta$ & $\alpha$ & $\beta$ & $\alpha$ & $\beta$ & $\alpha$ & $\beta$ & $\alpha$ & $\beta$ & $\alpha$ & $\beta$ & $\alpha$ & $\beta$ & $\alpha$ & $\beta$ & $)]-\frac{1}{2}[\frac{1}{5}($ & $\alpha$ & $\beta$ & $\alpha$ & $\beta$ & $\alpha$ & $\beta$ & $\alpha$ & $\beta$ & $\alpha$ & $\beta$ & $)]$\\
\hline
c  & 2 & & 1&1&1&1&1&1&1&1&1&1&1&1&1&1&1&1&1&1&1&1 &  & 1&1&1&1&1&1&1&1&1&1 &\\
o1 & 1 & & 1&0&1&0&1&0&0&1&1&0&1&0&0&1&1&0&0&1&0&1 &  & 1&0&1&0&1&0&1&0&0&1 &\\
o2 & 1 & & 1&0&1&0&0&1&0&1&1&0&0&1&1&0&0&1&1&0&1&0 &  & 1&0&1&0&1&0&0&1&1&0 &\\
o3 & 1 & & 1&0&0&1&0&1&1&0&0&1&1&0&0&1&1&0&1&0&1&0 &  & 1&0&1&0&0&1&1&0&1&0 &\\
o4 & 1 & & 0&1&0&1&1&0&1&0&1&0&0&1&1&0&1&0&0&1&1&0 &  & 1&0&0&1&1&0&1&0&1&0 &\\
o5 & 1 & & 0&1&1&0&1&0&1&0&0&1&1&0&1&0&0&1&1&0&0&1 &  & 0&1&1&0&1&0&1&0&1&0 &\\
\hline
\end{tabular}
\caption{ROKS operators for five unpaired electrons, doublet case.}
\label{taboperator5doublet}
\end{scriptsize}
%\end{rotate}
\end{table}

\section{Conclusions}

We have derived a general explicit energy expression for restricted open-shell 
Kohn-Sham theory which fulfills the sum rule for the energies.
If degeneracy is not prescribed and
the occupation pattern is interpreted as a symmetry, the energy
expression is also valid for the sum method by Ziegler, Rauk, and Baerends,
and, if the exact energy expectation values are used, for ROHF itself. 
In the latter case the energy expression
reduces to the exact energy of a single configuration \cite{Slater72}. \\
By inserting the Kohn-Sham expressions, we have derived ROKS operators
and have given explicit expressions for these operators for up to five electrons. \\
The energy expression and the operators constitute just part of the
restricted open-shell problem. For more than one open shell, only the
high-spin solution is easily obtained. For the low-spin
solutions of the ROKS equations, specific SCF algorithms turned out to be
useful in the case of two open shells. In future work we want to extend these
algorithms to more open shells.

\section{Acknowledgment}

This work was supported by the Deutsche Forschungsgemeinschaft:
SFB 486 'Manipulation von Materie auf der Nanometerskala', 
SFB 749 'Dynamik und Intermediate molekularer Transformationen', 
and the Nanosystems Initiative Munich (NIM). The authors thank
Sigrid Peyerimhoff and Jana Friedrichs for helpful comments and
David Coughtrie for reading the manuscript.

\end{document}